# Effect of aging-induced disorder on the quantum transport properties of atomically thin WTe$_2$

W. L. Liu[1,2], M. L. Chen[1,2], X. X. Li[1,2], S. Dubey[3], T. Xiong[1,2], Z. M. Dai[1,2], J. Yin[4], W. L. Guo[4], J. L. Ma[5], Y. N. Chen[5], J. Tan[1,2], D. Li[1,2], Z. H. Wang[1,2], W. Li[5*], V. Bouchiat[3], D. M. Sun[1,2], Z. Han[1,2*], and Z. D. Zhang[1,2]

[1] Shenyang National Laboratory for Materials Science, Institute of Metal Research (IMR), Chinese Academy of Sciences (CAS), 72 Wenhua Road, Shenyang 110016, China

[2] School of Material Science and Engineering, University of Science and Technology of China, Anhui 230026, China

[3] University of Grenoble Alpes, CNRS, Institut Néel, F-38000 Grenoble, France

[4] State Key Laboratory of Mechanics and Control of Mechanical Structures, Key Laboratory for Intelligent Nano Materials and Devices of the Ministry of Education and Institute of Nanoscience, Nanjing University of Aeronautics and Astronautics, Nanjing 210016, China and

[5] Institute for Advanced Study, Shenzhen University, Nanhai Avenue 3688, Shenzhen 518060, China

Atomically thin layers of transition-metal dicalcogenides (TMDCs) are often known to be metastable in the ambient atmosphere. Understanding the mechanism of degradation is essential for their future applications in nanoelectronics, and thus has attracted intensive interest recently. Here, we demonstrate a systematic study of atomically thin WTe$_2$ in its low temperature quantum electronic transport properties. Strikingly, while the temperature dependence of few layered WTe$_2$ showed clear metallic tendency in the fresh state, degraded devices first exhibited a re-entrant insulating behavior, and finally entered a fully insulating state. Correspondingly, a crossover from parabolic to linear magnetoresistance, and finally to weak anti-localization was seen. Real-time Raman scattering measurement, together with transmission electron microscopy studies done before and after air degradation of atomically thin WTe$_2$ further confirmed that the material gradually form amorphous islands. It thus leads to localized electronic states and explains the low temperature Coulomb gap observed in transport measurements. Our study reveals for the first time the correlation between the unusual magnetotransport and disorder in few-layered WTe$_2$, which is indispensable in providing guidance on its future devices application.

The recent finding of known largest magnetoresistance in WTe$_2$ [1] triggered numerous studies in this layered material [2], including pressure-driven superconductivity [3] and the predicted new type of Weyl semimetal state [4, 5]. Albeit a layered material, WTe$_2$ devices in the two-dimensional (2D) limit have been rarely reported, with its experimental investigations mostly restrained in the bulk regime. Unlike other transition-metal dichalcogenides (TMDCs), instead of 2H phase, Td-(also addressed as 1T'-) phase of bulk WTe$_2$ occupies the lower energy state, whose two nearly perfectly compensated electron and hole bands result in a large unsaturated classical magnetotransport, with a parabolic dependence of the applied magnetic field B [6]. Meanwhile, strong anisotropy was found in its bulk form, which gives rise to some exotic linear magnetoresistance in a specific measurement configuration [7]. Moreover, recent angle resolved photon electron spectroscopy (ARPES) studies showed subtleness of the band structure of bulk WTe$_2$ [8–11], intriguing possible peculiar electronic properties in the few layered scenario. In the 2D limit, electrostatic gate can, in principle, largely tune the carrier density, thus breaking down the electron-hole balance, leading to new opportunities.

Despite the fact that WTe$_2$ can be readily exfoliated against the weak interlayer van-der-Waals bonding, atomically thin WTe$_2$ is proven to rapidly age in ambient atmosphere like many of the TMDCs [12]. Together with the extinction of optical contrast and Raman signal after air exposure [13], ultra thin WTe$_2$ flakes were reported to exhibit, rather than the so-called semi-metal state, an insulating behavior as a function of number of layers, with the lack of a systematic analysis of the aging effect on the quantum transport [14]. Such air instability therefore hampers the further possibilities for nanoelectronic devices of, for example, few-layered WTe$_2$ field effect transistors [15]. However, so far, a quantitative understanding of the correlation between quantum transport behavior and the imposed disorder has been missing not only on atomically thin WTe$_2$, but also on other two dimensional electronic systems. To address this matter, it is of great importance to assess the detailed transport properties including IV characteristics of few-layered WTe$_2$ devices in the parameter space of temperature and magnetic field.

In this letter, we show thinning of bulk WTe$_2$ down to a few layers via mechanically exfoliating high quality crystal. Atomically thin flakes of WTe$_2$ were systematically studied via Raman and atomic force microscope (AFM) characteriz-

---

* Corresponding to: vitto.han@gmail.com, and wu.li@szu.edu.cn

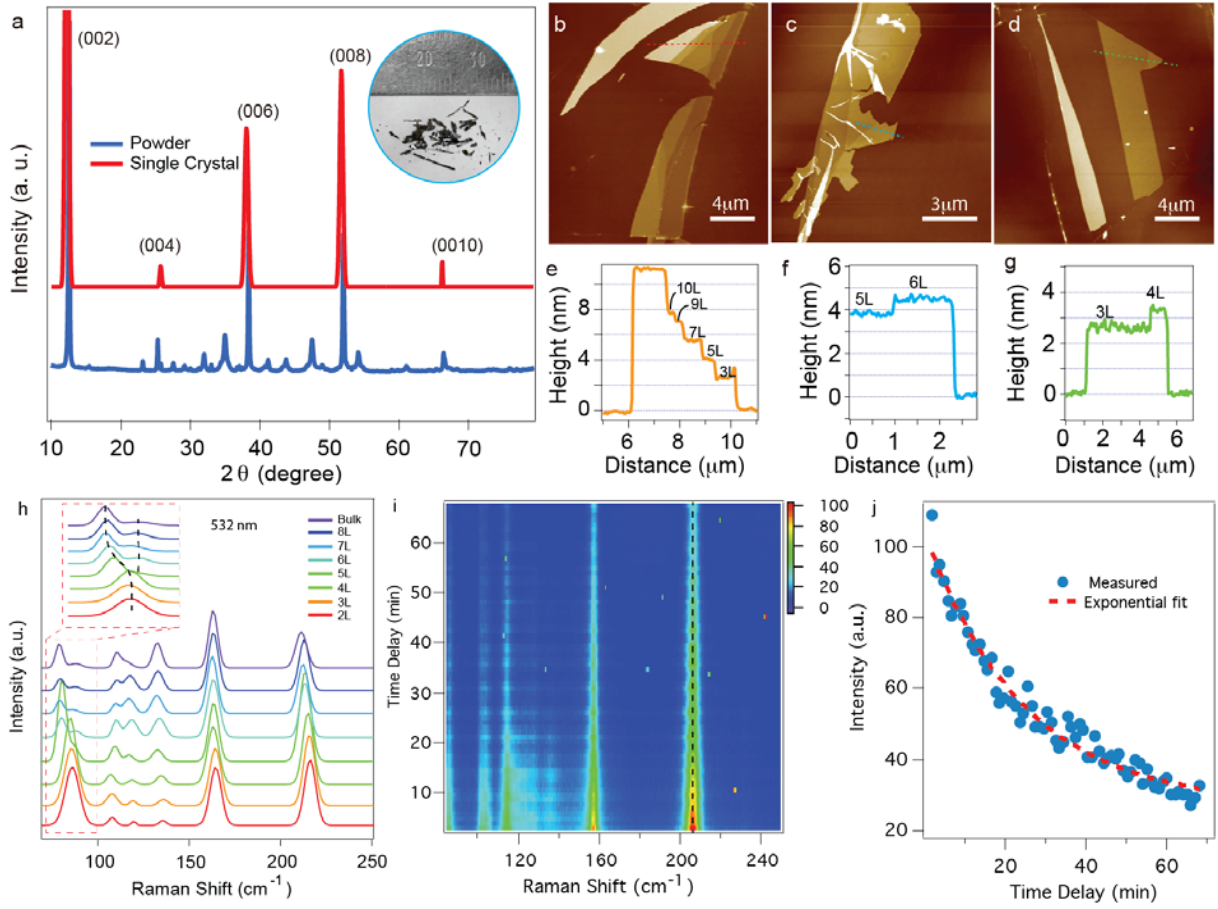

FIG. 1. (a) XRD pattern of Td-WTe$_2$. Red and blue are patterns for single crystal and powder samples, respectively. Inset in a) shows an optical image of the as-grown mm-sized crystal. (b)-(d) AFM images of typical WTe$_2$ flakes with different thickness. Height profiles along corresponding dashed lines are given in (e)-(g). (h) Green laser 523 nm Raman spectra for WTe$_2$ with different number of layers. Curves are fitted by single Lorentzian of each measurable peaks. Inset shows renormalized Raman characteristics at wavenumbers below 100 cm$^{-1}$. (i) Time evolution of Raman intensity for a four-layered WTe$_2$. (j) Peak intensity (highlighted by dashed line in (i)) as a function of time delay. Red dashed line gives the exponential fit.

-ations. We further fabricated the few-layered WTe$_2$ devices, and investigated their quantum electronic transport behavior with its evolution of aging. Strikingly, while the temperature dependence of few layered WTe$_2$ showed clear metallic tendency in the fresh state, degraded devices first exhibited a re-entrant insulating behavior, and finally entered a fully insulating state. Correspondingly, a crossover from parabolic to linear magnetoresistance, and finally to weak anti-localization (WAL) is observed. By investigating the current biased differential resistance of the few-layered WTe$_2$ devices, zero-bias $dV/dI$ peaks were observed, indicating a Coulomb gap due to electron-electron interaction. Transmission Electron Microscopy (TEM) studies before and after air degradation of atomically thin WTe$_2$ further suggested that the material gradually form amorphous islands, thus leading to localized electronic states. Our study reveals for the first time the correlation between the unusual magnetotransport and disorder in few-layered WTe$_2$, which provides significantly new understanding and extends the knowledge of the previous works on the issue of aging of TMDCs [12, 15].

Single crystal WTe$_2$ was prepared via the solid state reaction method. Raw material powders with stoichiometric ratio of W (purity 99.9 %) : Te (purity 99.99%) = 1 : 49 were mixed and kept at 1000 $^o$C for 8 hours. The mixture was then cooled at the rate of 2 $^o$C/h, followed by a centrifuge at 700 $^o$C. Figure 1a shows the X-Ray diffraction (XRD) pattern of powder (blue) and the exfoliated a-b plane (red) of the as-grown crystal shown in the inset. The highly oriented texture peaks of XRD reveal a single crystal of Td-WTe$_2$. We then applied the scotch tape method to exfoliate the bulk and deposited few-layered WTe$_2$ onto 285 nm thick silicon oxide grown on heavily doped silicon wafers. AFM images of such typical flakes with different number of layer are shown in Fig 1b-d. In the reported studies, WTe$_2$ layer thickness varies between 0.6 and 1.2 nm [13, 14, 16–18]. Here, we evaluated the thickness of a monolayer WTe$_2$ to be 0.79 nm, upon a statistics of more than 20 measured flakes. Such thickness profiles are given in Fig. 1e-g, cutting along the corresponding dashed lines in Fig.1 b-d. We notice that the probability of having mono-layer WTe$_2$ was rather low, we mostly had constant yields of above two layer WTe$_2$ thin flakes.

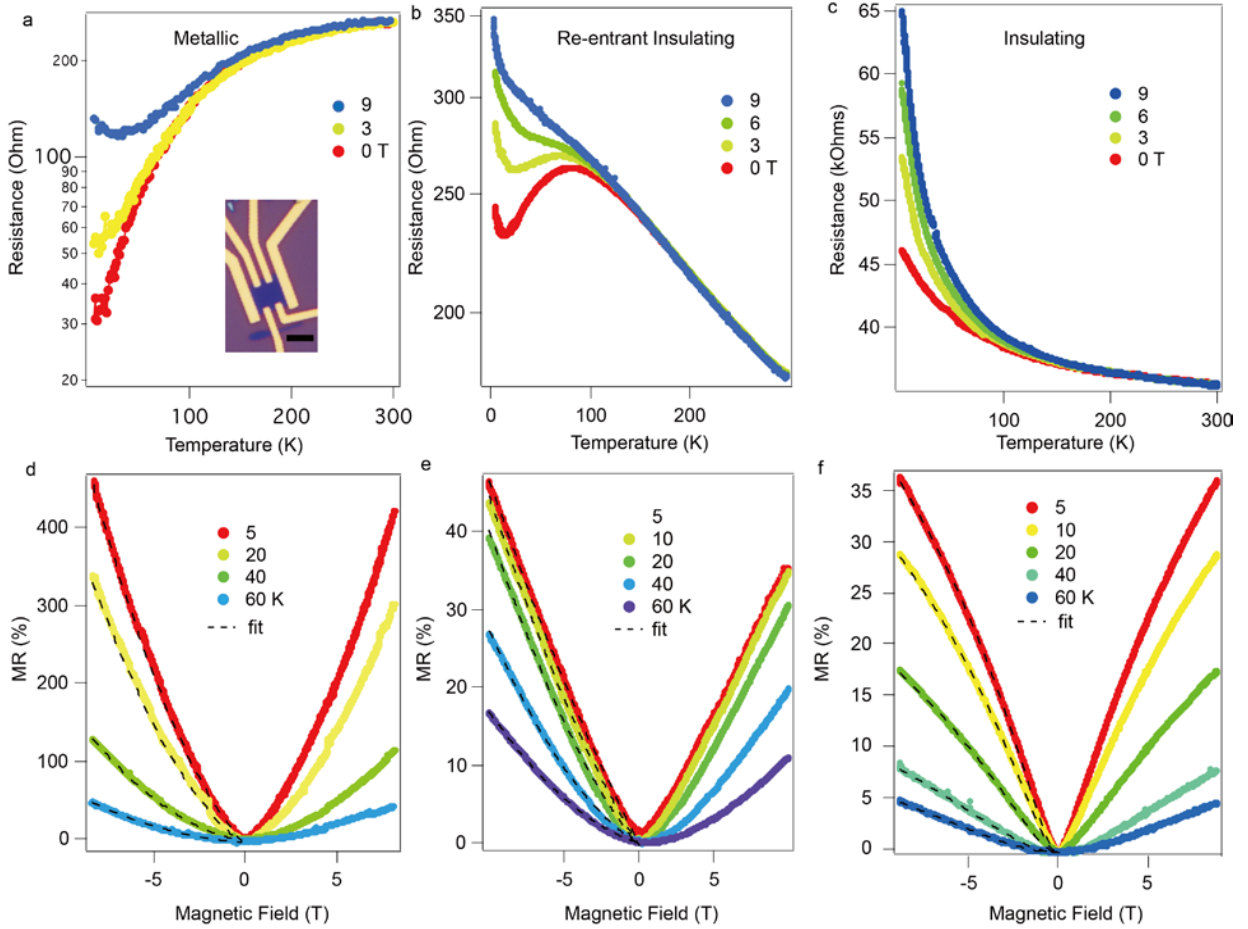

FIG. 2. (a)-(c) Resistance of 4-layered WTe$_2$ as a function of temperature at fixed magnetic field, and (d)-(f) Magnetoresistance of 4-layered WTe$_2$ at different temperatures. Inset in (a) shows the optical image of a typical measured device. Scale bar is 5 µm. Dashed line in (d)-(f) are fitted MR. (a)&(d) metallic device; (b)&(e) re-entrant insulating device; (c)&(f) insulating device.

Raman studies of WTe$_2$ were performed, in order to comprehend its lattice dynamics and the effects caused by air-degradation. Thin layers are exfoliated in ambient atmosphere for AFM and Raman measurements. Figure 1h displays the Raman spectroscopy of WTe$_2$ of 2 to 8 layers probed by a 532 nm wavelength laser. The bi-layer shows least peaks and lowest signal, with 6 distinct Raman bands observed around 86.7, 108.3, 120.0, 135.6, 164.2 and 215.3 cm$^{-1}$, respectively, in agreement of those reported in previous works [16–18]. All spectra are fitted via single Lorentzian and renormalized for visual clarity as seen in Fig. 1h. It is worth noting that the peak at wave number of about 86 cm$^{-1}$ splits into two bands above 4 layers, as highlighted in the dashed box in Fig. 1h. This anomaly was not reported before and may provide a new reference for distinguishing the number of layers in thin WTe$_2$.

Since ultra-thin WTe$_2$ is known to degrade rapidly due to air instability [13], it is of importance to have quantitative evaluation of its lifetime in ambient condition. Fig. 1i shows a time evolution of a 4-layered flake under 532 nm laser with 0.7 mW power. Clear decay of the peaks intensity with increasing time can be seen. Time evolution of the 215 cm$^{-1}$ peak is plotted in Fig. 1j. It is seen that the intensity as a function of time follows an exponential decay, which is fitted to a lifetime of $\tau_{air}$ = 62 min. On the opposite, thicker (more than 5 layers) flakes retain their Raman signal for days and even weeks, as they are screened by surface passivation of the top layer.

In the following, we will discuss the electron transport properties of few-layered WTe$_2$. A standard 4-probe method was used, while the magnetic field was applied perpendicularly to the exfoliated a-b plane of WTe$_2$ crystal. To start with, Fig. S1a shows the temperature dependence of bulk resistance (RT curves) at different magnetic fields. As can be seen, bulk WTe$_2$ has its symbolic "turn on" behavior, giving rise to the metal-to-insulator transition when subjected to magnetic field. Another perspective to present the curves is shown in Fig. S1b, i.e., magneto-resistance at different temperatures. All curves show parabolic B-dependence of the magnetoresistance signal. Indeed, here the magnetoresistace (MR) can be explained by the classical two band model, written as [6]:

$$\frac{\Delta R}{R} = \frac{\mu_e \mu_h (n\mu_e + p\mu_h)(p\mu_e + n\mu_h)B^2 - (p-n)^2 \mu_e^2 \mu_h^2 B^2}{(n\mu_e + p\mu_h)^2 + (p-n)^2 \mu_e^2 \mu_h^2 B^2} \quad (1)$$

where $R$ is the resistance, $n$, $p$, $\mu_e$ and $\mu_h$ are electron and hole densities and mobilities, $B$ is the magnetic field. In the case of fully balanced electron and holes, it reduces to a quadratic form $\frac{\Delta R}{R} = \mu_e \mu_h B^2$. Curves in Fig. S1b follow closely this quadratic trend, indicating the well-known electron-hole compensation in the bulk.

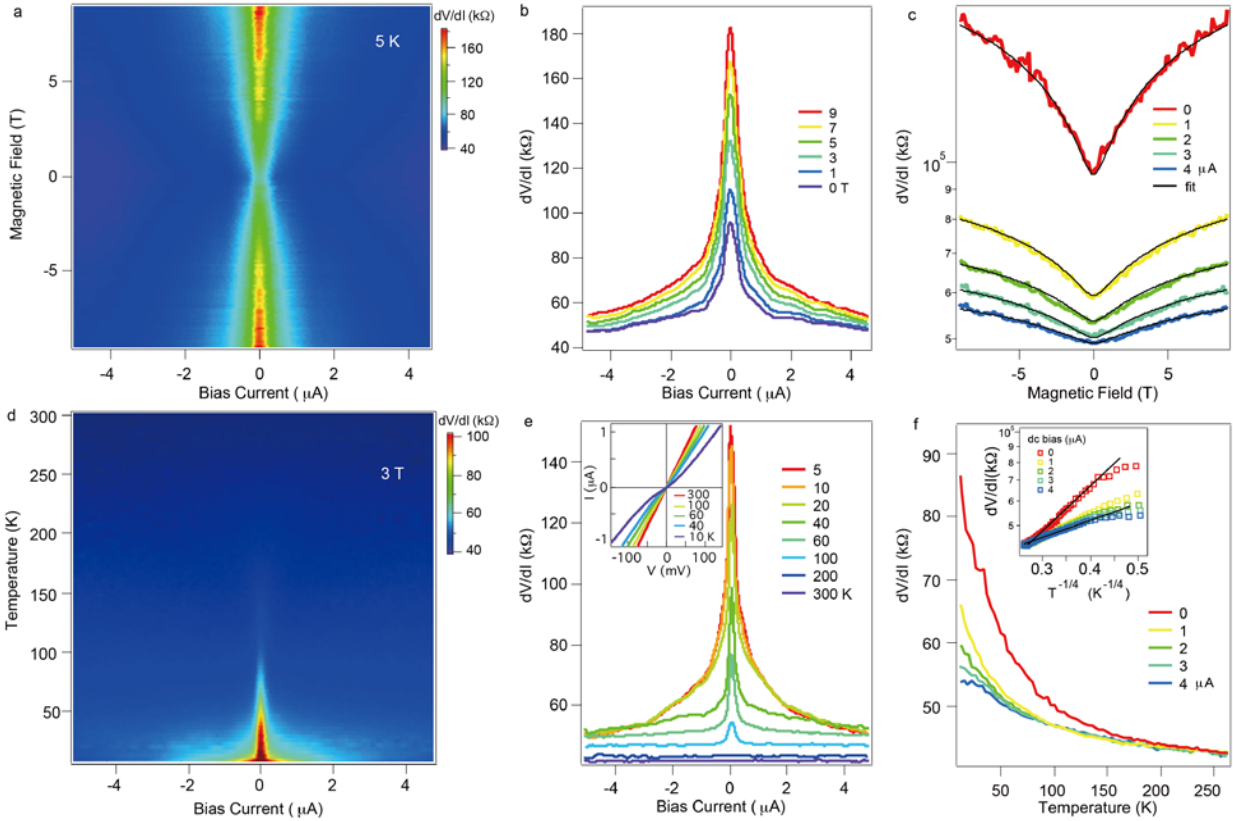

FIG. 3. (a) Differential resistance of the sample in Fig. 2c, measured as a function of bias current and magnetic field at 5 K temperature, with its line cuts along fixed magnetic field and bias current shown in (b) and (c), respectively. Solid lines in (c) are fitted by Equation (2). (d) Differential resistance measured as a function of bias current and temperature at 3 T magnetic field, with its line cuts along fixed temperature and bias current shown in (e) and (f), respectively. Inset in (e) shows reconstructed IV characteristics by integrating the dV/dI curves in (e). Inset in (f) shows a semi-log plot of $dV/dI$ against $T^{-1/4}$. Linear dependence can be seen at high temperature ranges.

We now come to the fabrication of few-layered WTe$_2$ devices. Multiple samples of few-layered WTe$_2$ with different thickness were contacted by Ti (5 nm)/Au (60 nm) electrodes via standard e-beam lithography. As discussed in the previous section, ultra-thin WTe$_2$ flakes suffer from air instability. To minimize exposure to air and hence to measure in a "fresh" state the devices, a resist (PMMA) layer was immediately (after a few minutes of exposure) spun on few layered WTe$_2$ flakes after exfoliation. Identifying under optical microscope and lithography are then performed with the PMMA protection. Electrodes metallization were done followed by rapid re-spun of PMMA resist and a second exposure was made to open windows on the bonding pads, with the rest of PMMA resist kept throughout the measuring process. Figure 2a shows the 4-probe RT curves under different magnetic fields of a typical 4-layered WTe$_2$ device. Thanks to the capping layer of PMMA, the device showed clear metallic state, as opposed to the insulating behavior reported elsewhere [14]. Interestingly, magnetoresistance at different temperatures of the 4-layered device does not follow exactly the parabolic B-dependence, as shown in Fig. 2d. Instead, an additional correction of linear magnetoresistance is needed to fit the data (dashed lines in Fig. 2d). Thus the fit is given by $\frac{\Delta R}{R} = \alpha B^2 + \beta B$, where $\alpha$ and $\beta$ are fitting parameters.

To further investigate the influence of imposed disorder in few layered WTe$_2$, we measured a 4-layer sample, which was exposed to air for 1 week without the protecting PMMA layer. Strikingly, the device with fairly disordered state showed a re-entrant insulating behavior (Fig. 2b), with a metallic intermediate region appearing in the temperature range of 10-80 K at zero magnetic field. The intermediate metallic state was then totally suppressed by cooling with a magnetic field above 6 T. The reentrant insulating behavior has been already reported in other systems such as amorphous superconducting film [19], and ultra-thin manganite compounds [20], and was attributed to localization or electronic phase separation, respectively. In the present system, we speculate that the onset of metallic conduction of WTe$_2$ appears only in the intermediate temperature range, and electrons stay localized both at room temperature and at the lowest temperature region. Magnetoresistance of this device (exposed to air for a week without PMMA protection) is shown in Fig. 2e, which gives much more pronounced characteristics of linear $B$-dependence compared to the fresh sample shown in Fig. 2d. The increased disorder with concomitantly enhanced linear MR, together with persisting quadratic $B$-dependence at low field, point towards a possible disorder-driven linear MR [21–23]. It is interesting to note that the disorder induced magneto-resistance change here (linear MR to weak anti-localization) is quite similar to another layered 2D telluride system, Bi$_2$Te$_3$, as reported previously [24, 25]. From the applied physics point of view, a linear non-saturating magnetoresistance is highly desired for the design of magnetic sensors. Few layered WTe$_2$ therefore seems to be one of the promising candidates.

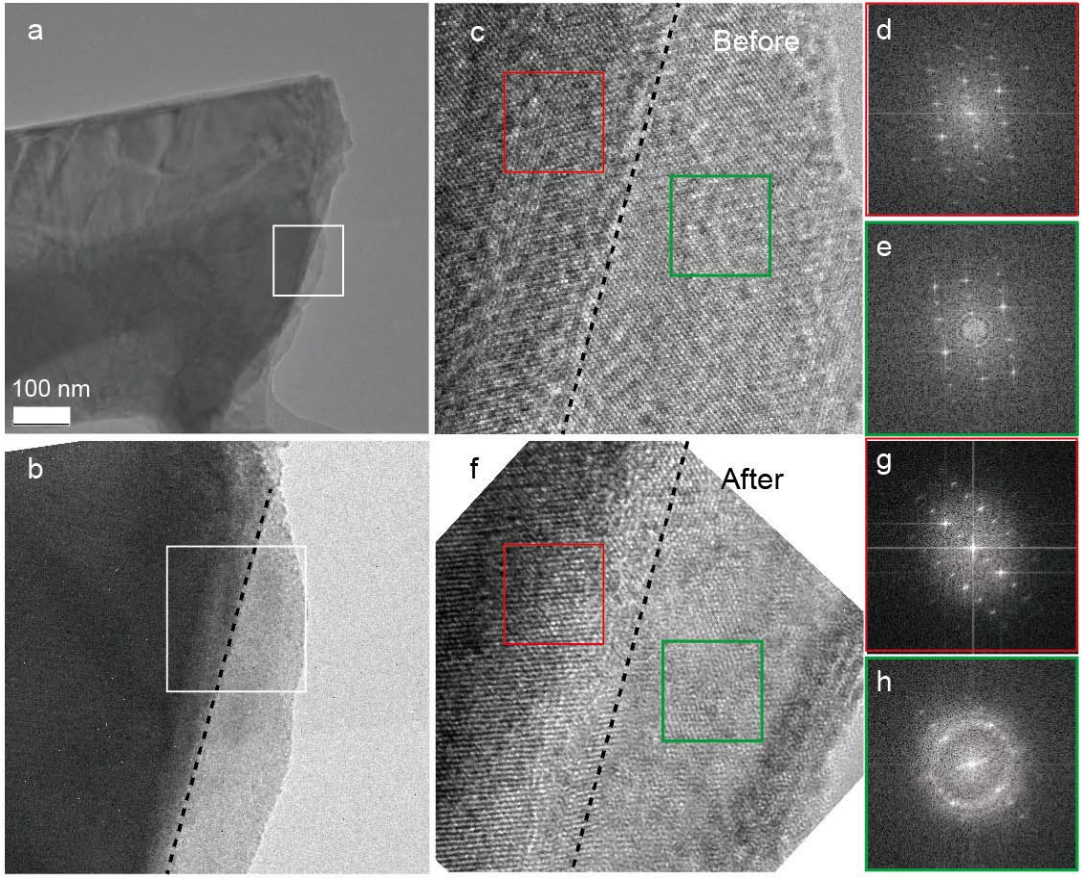

FIG. 4. (a) TEM image of WTe$_2$ flake, whose white boxed area is shown in (b). Boxed area in (b) is then zoomed for high resolution TEM measurement. (c) and (f) are images taken before and after thermal degradation of the flake in air, respectively, with their boxed area performed FFT, as shown in (d)-(e) and (g)-(h). Dashed lines in (b)-(f) highlights the boundary between thick and thin WTe$_2$. Notice (f) is rotated to align with (b) & (c) for visual clarity.

As shown in Fig. 2c, when the flakes are strongly degraded (either by long exposure to air or by heating to above 200 $^\circ$C for a few min in air) WTe$_2$ devices with number of layers between 3 and 8 all show exponentially increasing resistance upon cooling, i.e., a complete insulating state. Fig. 2f shows the magnetoresistance of the insulating 4-layered device at different temperatures, which drastically differs from the metallic and re-entrant insulating states. A weak anti-localization characteristic was observed, which is described by the Hikami-Larkin-Nagaoka theory [26]:

$$\Delta\sigma \propto \Psi(0.5 + \frac{B_\Phi}{|B|}) - \ln(\frac{B_\Phi}{|B|}) \qquad (2)$$

where $\Psi$ is the digamma function, and $B_\Phi = \frac{\hbar}{4eD}\tau_\phi^{-1}$, with $D$ the diffusion constant and $\tau_\phi$ the electron phase coherent time. In equation 2, we neglected other spin-orbit coupling terms which play negligible role in the fitting process. WAL observed in 2D electronic systems, including topological insulators and graphene, are often explained by a π Berry phase captured by electrons through closed trajectory [27, 28]. Recent studies in 3D Dirac semimetal, 3D Weyl semimetal, as well as chemical vapor deposited TMDCs also showed WAL phenomenon at relatively low field range [29, 30].

In mesoscopic devices, differential resistance *dV/dI* as a function of bias current is a useful tool to analyze the transport behavior (see Methods section). Figure 3a shows a color map of *dV/dI* at 5 K temperature in the dc bias current and magnetic field space for the same device measured in Fig. 2c and 2f (multiple strongly degraded samples with different number of layers all showed similar *dV/dI* behavior). Line cuts along fixed B show a strong zero-bias resistance peak, as shown in Fig. 3b. This zero-bias anomaly is a characteristic of Coulomb gap induced by the local electron charging effect. In a disordered system, electron transport occurs via variable range hopping, while Coulomb interactions between different hopping sites are important and lead to a low-bias barrier at low temperature. Line cuts along fixed bias current of Fig. 3a are also plotted, as shown in Fig. 3c. It is seen that at low bias, the magnetoresistance show highest absolute values, while all cases, including the zero-biased, can be well fitted by the WAL using equation 2.

An important picture to evaluate the Coulomb gap in the studied WTe$_2$ few layers is the temperature dependence of the *dV/dI* vs bias current. Fig. 3d shows a representative color map of such measurement under 3T magnetic field. It is clearly observed in the line cuts along fixed temperature (Fig. 3e) that the *dV/dI* curves above 200 K are completely flat, while the zero-bias anomaly starts to show up below 200 K, and increase with lowering the temperature as expected for a typical Coulomb gap anomaly. To further investigate the output characteristics, we plotted in *IV* curves by integration of differential resistance, shown in the inset of Fig. 3e. One can see that in the high temperature range, *IV* curves are linear, corresponding to Ohmic transport. At base temperature of 5 K, *IV* develops into a semiconductor-like non-linear state. Finally, we examine *RT* curves at different bias current, as shown in Fig. 3f, whose high temperature range can be well explained by the 3D Mott's hopping law $R \propto \exp[(T_0/T)^{1/4}]$ (inset of Fig. 3f), while divergence are found at low temperatures. One possibility to explain the Mott's hopping behavior is the extremely small electron scattering distance that is even smaller than the few-layer thickness, and the surface charging effect is the main cause of the observed Coulomb blockade phenomenon. Surprisingly, it is noticed that even for fresh metallic WTe$_2$ devices which were protected to minimize air exposure, the Coulomb gap already started to develop below 60 K, as shown in Fig. S2. Interestingly, recent study suggested such gap to be of a possible quantum spin Hall origin [31]. Our above analyses suggest that this nonlinear *IV* characteristic is mainly caused by disorder, as indicated by the correlated increase of Coulomb gap.

In order to reveal the microscopic origination of the observed unusual magnetotransport, we performed TEM analysis of few layered WTe$_2$ before and after heat treatment. As shown in Fig. 4a and b, an ultra-thin area of the freshly prepared specimen (see Methods section) was located. High-resolution TEM image shown in Fig. 4c indicates good crystalline structure on both sides of the thick and thin areas separated by the black dashed line. Fast Fourier transform (FFT) of boxed areas in Fig. 4d-e shows almost identical lattice symmetry on both sides, which proves the pristine lattice in atomically thin WTe$_2$ layers. By taking out the TEM specimen and heating in air at 200 $^o$C for 5 min, the thin WTe$_2$ become significantly degraded. The same flake was relocated after the heat treatment, as given in Fig. 4f. It can be seen that on the left side (thick region), lattice are preserved as its fresh state. However, on the right side of the dashed line (thin region), the lattice become much degraded as expected, which exhibits typical amorphous features, with its FFT in Fig. 4h presenting the ring-shape compared to the ones in Fig. 4g and d-e. By examining the thin part of WTe$_2$ in Fig. 4f, it is found that WTe$_2$ crystalline clusters of about 10×10 atoms are embedded in the amorphous matrix when strongly degraded, agree with the hopping-like localized behavior in transport measurements. Moreover, the strongly degraded WTe$_2$ thin layers lost most of its optical contrast, while AFM scan shows almost the same height (Fig. S3), consistent with the scenario of becoming amorphous as observed by TEM.

Indeed, when plotted as differential conductance *dI/dV* against bias voltage, the Coulomb gap Δ in strongly degraded sample can be estimated to be at the order of few tens of meV at 5K (Fig. S4a). The same plot of "metallic" sample in Fig. S4b exhibits also such gap below 60 K, with however much lower magnitudes of a few tens of $\mu$eV. The rather low mobility of about 74 cm$^2$V$^{-1}$s$^{-1}$ (Fig. S5) extracted in such "metallic" WTe$_2$ field effect transistor may be already limited by the degradation.

As a summary, in few layered WTe$_2$ devices, we found a transition from the metallic state to a re-entrant insulating state, and finally to full insulating state that is correlated with increasing disorder. Correspondingly, a crossover from semi-metallic to linear magnetoresistance and finally to weak anti-localization is observed. Systematic studies by Raman, AFM, TEM, and differential resistance measurements have provided for the first time microscopic understanding of air instability of atomically thin WTe$_2$. It is thus expected that by reducing surface degradation, such as the recently developed boron nitride (h-BN) encapsulation technique [32, 33], the pristine quality of 2D WTe$_2$ may be preserved and may give rise to higher electron mobility, leading to WTe$_2$ transistors with better performances, as well as profound physics such as the seek for experimental proof of type-II Weyl fermions.

## METHODS

Raman and AFM measurement were carried out using a Renishaw and the Bruker icon system, respectively. 532 nm wave length and 0.5 mW power were applied in this study. To obtain the as fresh as possible ultra-thin $WTe_2$ devices, we spin coated a layer of PMMA resist immediately after exfoliation. Degraded samples are then studied without PMMA capping layer. We found that heating on a hot-plate above 200 $^o$C gives similar result as long air exposure (i.e., by keeping the sample in air for a few weeks or even month). Measurements carried out in Fig. 2 were using excitation of 5 µA ac current from Stanford SR830 lock-in with a bias resistor. While data in Fig. 3 were using a four probe dc bias configuration, added by an ac lock-in signal to obtain the $dV/dI$ curves. Lock-in excitation of 50 nA up to 5 µA were applied to add on the dc current sweep from a Keithley 2400 with a bias resistor. A sum of current was then filtered and amplified using a band pass pre-amplifier.

TEM specimens were prepared by transferring carbon grid onto freshly exfoliated $WTe_2$ flakes on a $SiO_2$ wafer. A droplet of iso-propanol was applied onto the grid and let dry in air before peeling off the TEM grid. TEM observation was carried out using an FEI Tecnai-F20 system. The same few layered $WTe_2$ flake (before and after degradation) was then observed under a 200 keV electron beam.

## ACKNOWLEDGEMENT

This work is supported by the National Natural Science Foundation of China with Grant 51522104 and 11504385. D.M. Sun thanks the National Natural Science Foundation of China with grant 51272256, 61422406, and 61574143. D. Li acknowledges the National Natural Science Foundation of China with grant 51371175. Z.D. Zhang acknowledges supports from the National Natural Science Foundation of China with grant 51331006 and the Chinese Academy of Science under the project KJZD-EW-M05-3. V. Bouchiat acknowledges support from the EU Graphene Flagship program, and the 2D-transformers program of Agence Nationale de la Recherche (ANR).

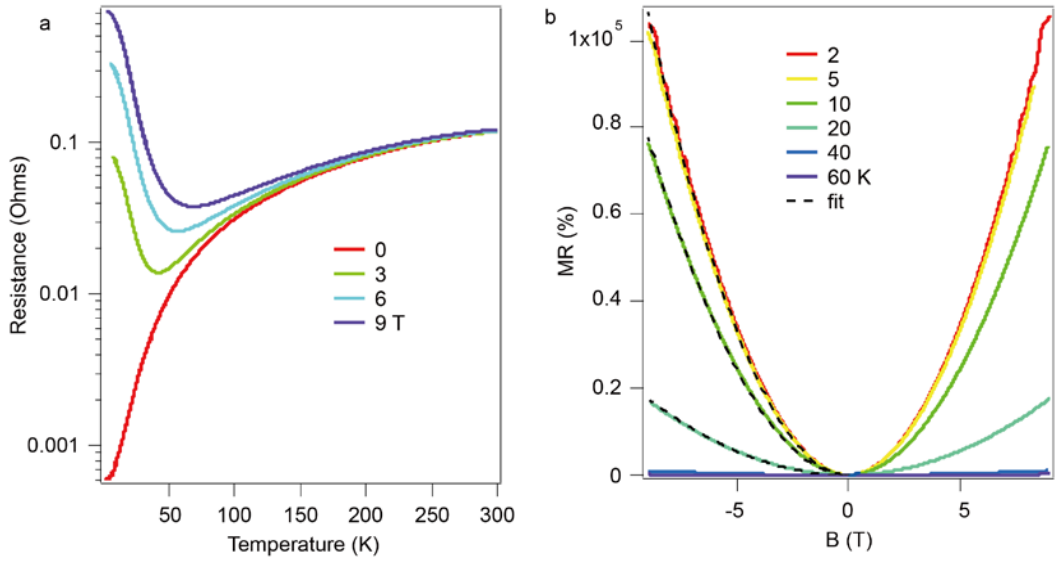

Fig.S. 1. (a) Four-probe bulk resistance as a function of temperature at fixed magnetic field, (b) Bulk magnetoresistance at different temperatures. Dashed lines in (b) are fitting using Equation (1) in the main text.

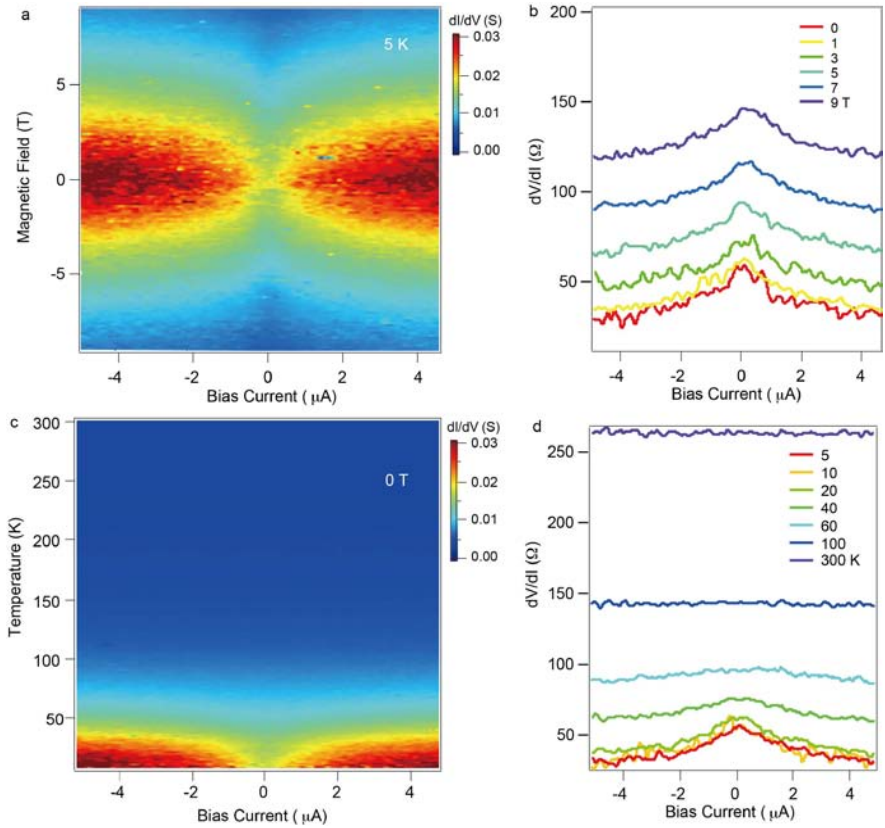

Fig.S. 2. (a) Differential conductance of the metallic sample (in main text Fig. 2a) as a function of bias current and magnetic field at 5 K temperature, with its line cuts along fixed magnetic field shown in (b)). (c) Differential conductance of the same device measured as a function of bias current and temperature at 0 T magnetic field, with its line cuts along fixed temperature shown in (d).

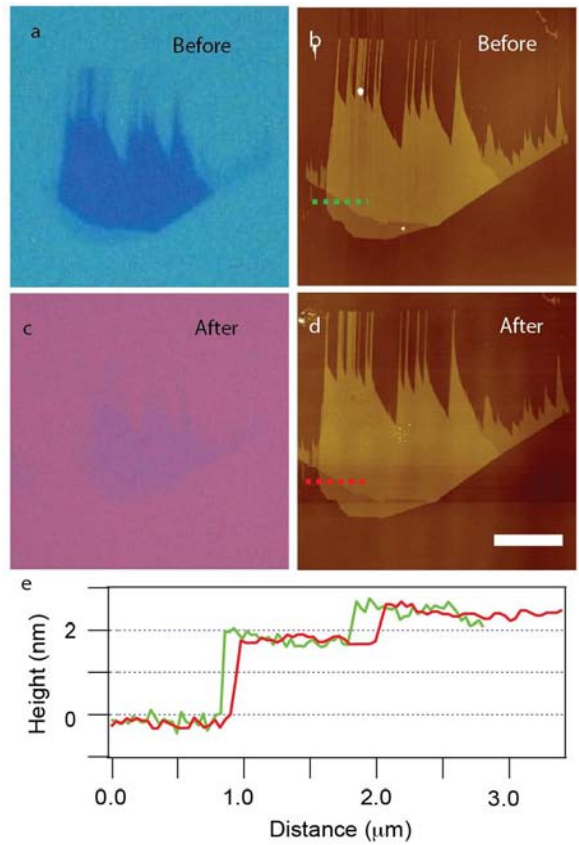

Fig.S. 3. (a)-(b), and (c)-(d) the optical and AFM images of the same flake before and after heated to 300 °C in air. The flake lost most of its optical contrast after heating. However, AFM scan in (d) shows nearly same height morphology as its pristine state in (b), as indicated in the line profiles in (e). Scale bar is 4 µm.

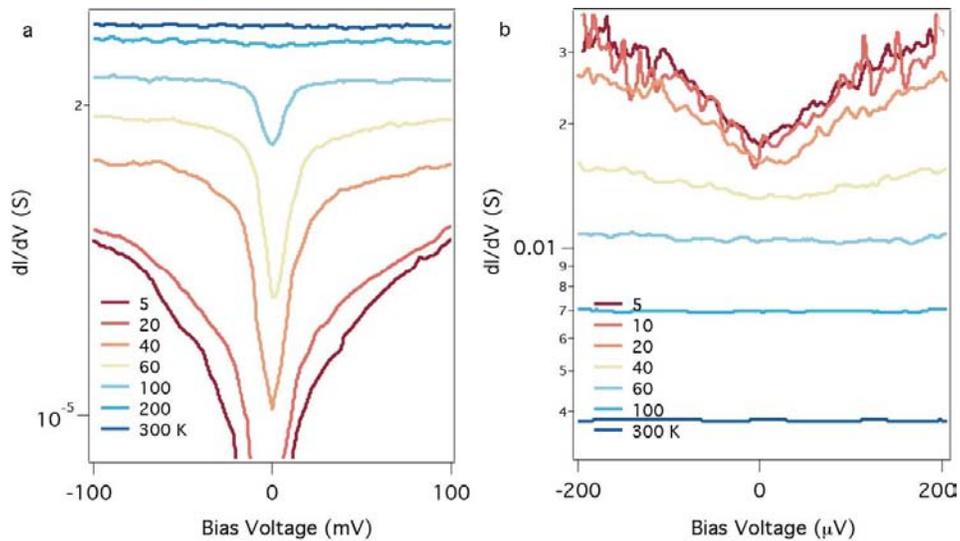

Fig.S. 4. (a) Differential conductance *dI/dV* plotted against bias voltage for the same data in Fig. 3e in the main text. (b) Differential conductance *dI/dV* plotted against bias voltage for the same data in Fig. S2d.

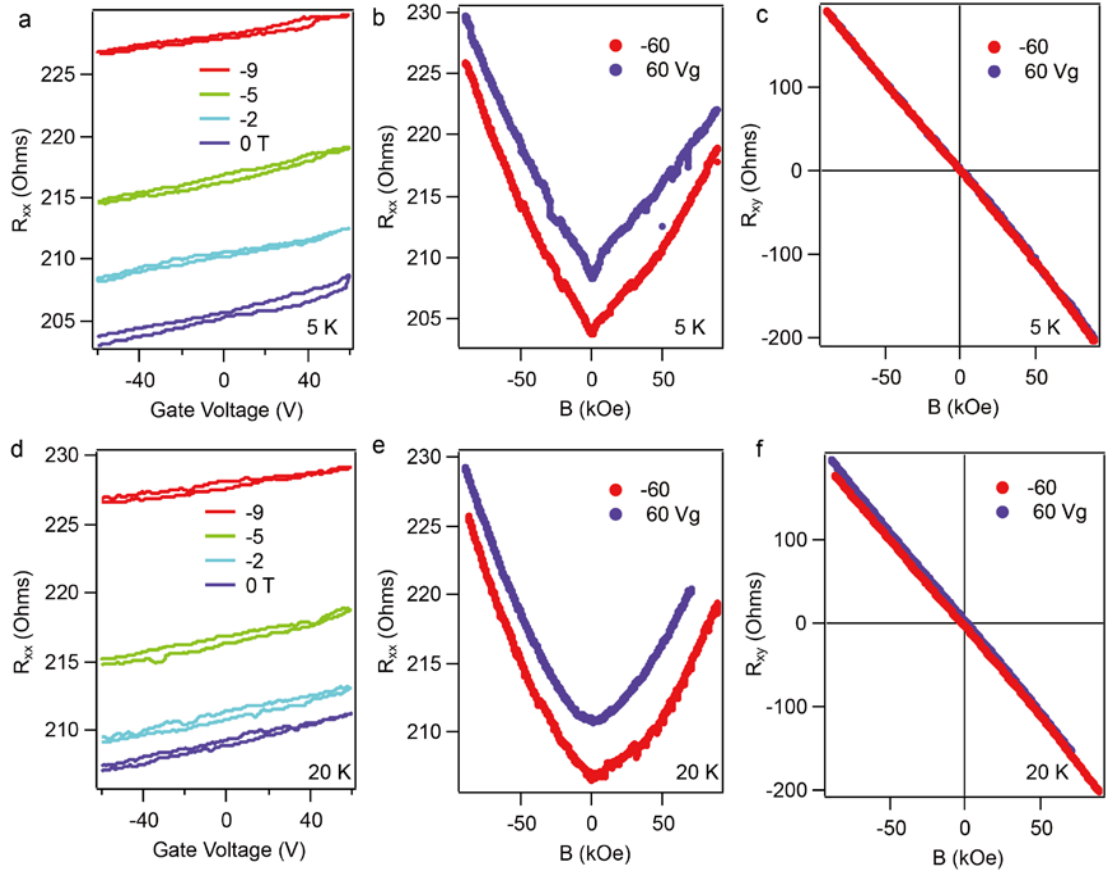

Fig.S. 5. (a) Field effect of the "metallic device at different magnetic field. (b) Magnetoresistance measured at -60 and 60 V gate voltages, with their corresponding transverse resistance $R_{xy}$ plotted in (c). (d)-(f) are same measurements at 20 K, compared to those measured at 5 K in (a)-(c). For square sample, by using the capacitive coupling model, electron mobility $\mu$ satisfies the relation $\sigma = ne\mu$, where $\sigma$ is the conductivity, n the carrier density, e electron charge, and $ne = CV_g$, with C the capacitance of dielectric layer, $V_g$ the gate voltage. When in the case of not knowing the doping level, it can be simplified to $\mu=(1/C)(d\sigma/dV_g)$. Mobility can therefore be evaluated from the slope of $\sigma$-$V_g$ plot. We thus estimate the in the measured device to be around 74 $cm^2V^{-1}s^{-1}$.